\documentclass[12pt,letterpaper]{iopart}
\usepackage{amssymb}
\usepackage{graphicx}
\usepackage{dcolumn}
\usepackage{bm}

\begin{document}
\title[]{The suppression of hidden order and onset of ferromagnetism in URu$_2$Si$_2$ via Re substitution}
\author{N P Butch\footnote{Present address: University of Maryland, College Park, MD 20742} and M B Maple}
\address{Department of Physics and Institute for Pure and Applied Physical Sciences, University of California, San Diego, La Jolla, CA 92093}

\date{\today}

\begin{abstract}
Substitution of Re for Ru in the heavy fermion compound URu$_2$Si$_2$ suppresses the hidden order transition and gives rise to ferromagnetism at higher concentrations.  The hidden order transition of URu$_{2-x}$Re$_x$Si$_2$, tracked via specific heat and electrical resistivity measurements, decreases in temperature and broadens, and is no longer observed for $x>0.1$.  A critical scaling analysis of the bulk magnetization indicates that the ferromagnetic ordering temperature and ordered moment are suppressed continuously towards zero at a critical concentration of $x \approx 0.15$, accompanied by the additional suppression of the critical exponents $\gamma$ and $\delta -1$ towards zero.  This unusual trend appears to reflect the underlying interplay between Kondo and ferromagnetic interactions, and perhaps the proximity of the hidden order phase.

\end{abstract}

\maketitle
For well over two decades, the identity of the ordered phase found in URu$_2$Si$_2$ at temperatures below 17~K has eluded researchers. This hidden order (HO) phase coexists with a heavy fermion state, but yields to a superconducting ground state below 1.5~K. In order to develop a better understanding of the HO phase, URu$_2$Si$_2$ has been extensively studied by various techniques, and these efforts have uncovered fascinating behavior under magnetic field, applied pressure, and chemical substitution. We describe the novel phase changes induced by Re substitution for Ru, including the suppression of the HO transition, the nearby emergence of ferromagnetic (FM) order, and the unique critical behavior of this phase.

Chemical substitution studies have been used primarily to investigate the robustness of the HO phase, and substitutions have been made for all three elements in URu$_2$Si$_2$.  Replacing U with Th or rare earths quickly suppresses the HO transition temperature \cite{Torre92,Park95}, emphasizing the critical role of the U \emph{f}-electrons to the HO phase.  In contrast, substitution for Si has no dramatic effects \cite{Park94}. Most of the neighboring elements of Ru: Mn, Tc, Th, Re, Os, Rh, and Ir have been substituted onto the Ru site, and in all cases except Os, the HO transition is suppressed by about 5\% substituent concentration \cite{Dalichaouch89,Dalichaouch90a,Dalichaouch90b}.

For a few elements, the phase diagrams at higher concentrations have been studied.  In the Rh case, in addition to the HO phase, there are two other nominally antiferromagnetic (AFM) phases, which are separated by nonmagnetic regions \cite{Miyako91}.  In contrast to the HO, these AFM phases are characterized by a larger moment \cite{Miyako92,Kohori92}, and a complex magnetic structure exists at intermediate Rh concentrations \cite{Kawarazaki94}.  A closer examination of the HO phase found that for almost half of the range of Rh concentration over which the HO phase exists, the order is actually AFM \cite{Yokoyama04}.

Substituting into URu$_2$Si$_2$ with Re, Tc, or Mn leads instead to ferromagnetic (FM) order ~\cite{Dalichaouch90b}, whose discovery constituted the first experimental realization of a heavy fermion FM instability.   In URu$_{2-x}$Re$_x$Si$_2$, a maximum $T_{C} \approx 40$~K is found at $x=0.8$, near the concentration limit below which the material remains single phase.  The itinerant FM order is characterized by a small moment, with no anomalies in electrical resistivity or specific heat at $T_{C}$. Across the Re-substitution series, the heavy fermion state is reflected in an enhanced  electronic contribution to the specific heat \cite{Dalichaouch89} and a narrow Drude peak in optical conductivity at low temperatures \cite{Thieme95}.  The long range nature of the FM phase has been confirmed by neutron scattering  \cite{Torikachvili92} and  $^{29}$Si NMR \cite{Kohori93}.  Within the FM phase, non-Fermi liquid (NFL) behavior is observed in the low-temperature bulk properties \cite{Bauer05} and in energy-temperature scaling of the dynamic magnetic susceptibility \cite{Krishnamurthy08}. The origin of the NFL effects remains an open question, partly because it has been difficult to conclusively define the FM phase boundary.  As only polycrystalline samples have been studied so far, it is also necessary to exclude orientational averaging or grain boundary effects as causes for these unusual properties.

Single crystals of URu$_{2-x}$Re$_{x}$Si$_{2}$ were synthesized via the Czochralski technique in a tri-arc furnace, and subsequently annealed in Ar. Phase homogeneity was verified by powder x-ray diffraction.  Samples  were oriented via the back-reflection Laue method and spark cut and sanded to size. Electrical resistivity measurements as a function of temperature  $\rho(T)$ were performed in a liquid $^4$He cryostat. Measurements of magnetization in field $M(H)$ were made using a Quantum Design Magnetic Properties Measurement System (MPMS).  Measurements of the specific heat $C$ were performed in a $^3$He refrigerator utilizing a semi-adiabatic heat pulse technique.

\begin{figure}[tbp]
    \begin{center}
    \rotatebox{0}{\includegraphics[width=8.5cm]{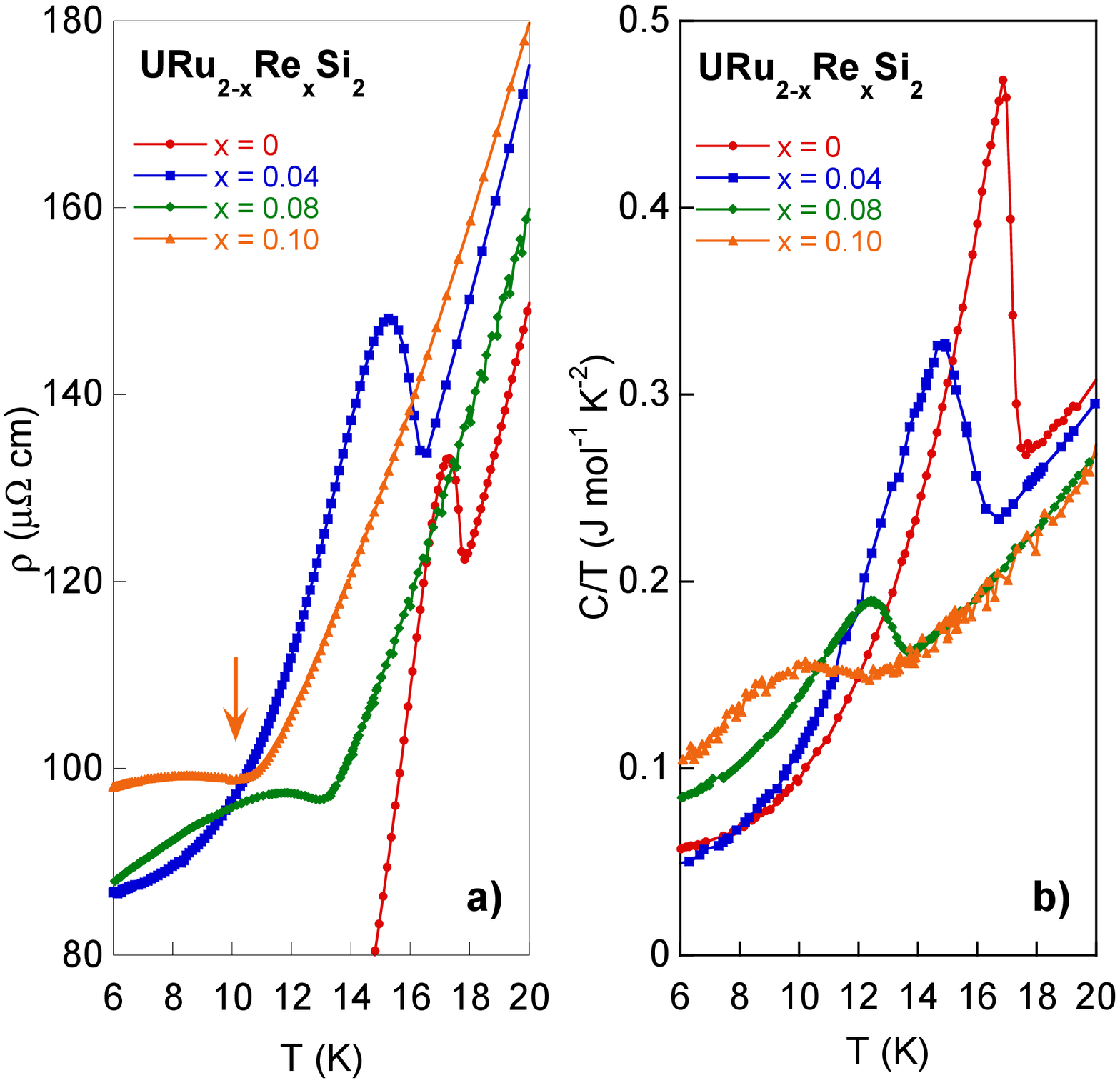}}
    \end{center}
    \caption{Electrical resistivity and specific heat anomalies associated with the hidden order transition in URu$_{2-x}$Re$_x$Si$_2$. a) The transition temperature $T_0$ decreases with increasing $x$, as the height of the anomaly in $\rho(T)$ shrinks and the width broadens.  The arrow points to the minimum in $\rho(T)$ for $x=0.10$. b) The jump in $C/T$ at $T_0$ decreases in magnitude as $T_0$ decreases with increasing $x$.}
    \label{HO}
\end{figure}

As seen in Fig.~\ref{HO}, in the vicinity of the HO transition, the $\rho(T)$ data exhibit a small peak, qualitatively similar to that seen in elemental Cr at the ordering transition into a spin density wave. The transition into the HO phase is marked by a large specific heat anomaly, which is related to the incomplete gapping of the Fermi surface in URu$_2$Si$_2$ \cite{Maple86,Behnia05,Sharma06}.  The general shape of these anomalies persists despite a suppression of the ordering temperature as $x$ increases, suggesting that the transition into the HO state always involves a Fermi surface instability.  As Re concentration increases, the anomalies shrink in height and broaden in width, becoming more poorly defined near $x=0.10$. For $x=0.12$, no HO anomaly can be identified in $\rho(T)$ or $C(T)$.

\begin{figure}[tbp]
    \begin{center}
    \rotatebox{0}{\includegraphics[width=8.5cm]{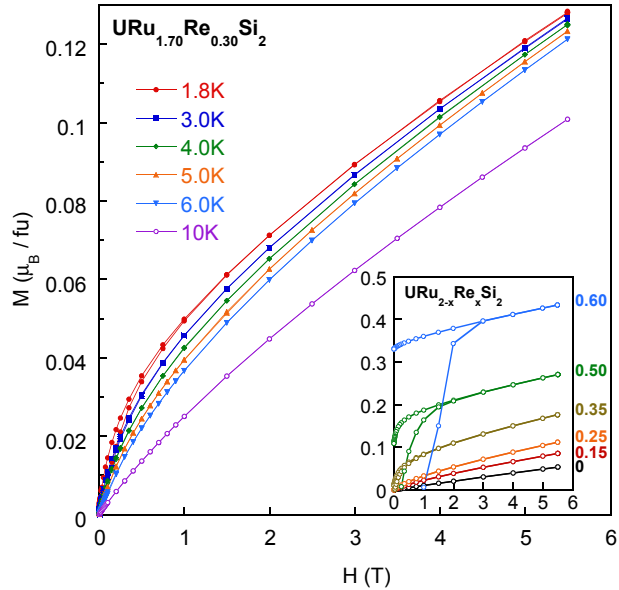}}
    \end{center}
    \caption{Magnetization isotherms for URu$_{1.70}$Re$_{0.30}$Si$_2$. Inset: a direct comparison of low-temperature
    data for different Re concentrations. Data for $x=0.60$ were measured at 1.8~K, while the other data were taken at 2.0~K.}
    \label{MH}
\end{figure}

Isotherms of $M(H)$ are shown in Fig.~\ref{MH} for URu$_{1.70}$Re$_{0.30}$Si$_{2}$, which orders ferromagnetically below approximately 5~K. As the Re concentration increases from $x=0.20$ to $x=0.60$, the magnitude of the magnetization increases, the isotherms become more sharply curved and hysteresis develops, which is clearly shown in the inset of Fig.~\ref{MH}.  Although the parent compound URu$_{2}$Si$_{2}$ exhibits linear $M(H)$ in this field range, in URu$_{2-x}$Re$_{x}$Si$_{2}$, the isotherms exhibit gentle curvature, which does not saturate.   Despite the dramatic increase of the low-$T$ magnetization with Re concentration, even at $x=0.60$, well within the FM state, the moment is a small fraction of the high-temperature paramagnetic effective moment of approximately 3.8~$\mu_\mathrm{B}$.

\begin{figure}[tbp]
    \begin{center}
    \rotatebox{0}{\includegraphics[width=8.5cm]{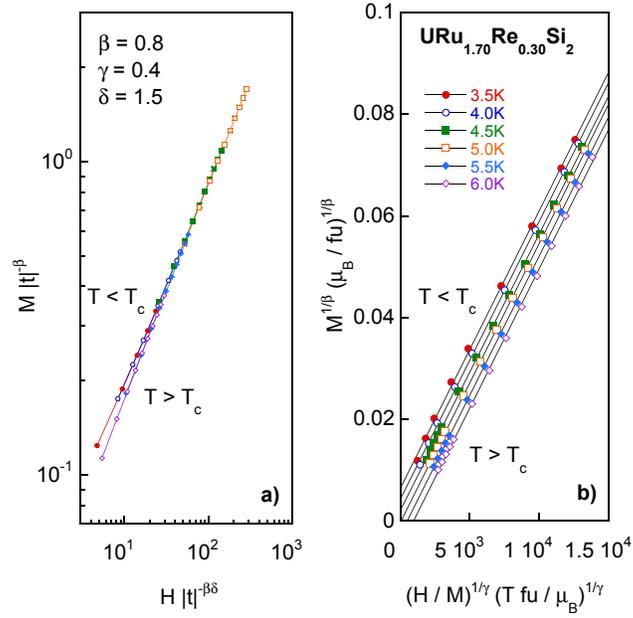}}
    \end{center}
    \caption{Arrott-Noakes scaling law and modified Arrott plots for URu$_{1.70}$Re$_{0.30}$Si$_{2}$. a) In the scaling plots, $|M|/|t|^{\beta}$ vs $H/|t|^{\delta\beta}$, the $M(H)$ data collapse onto two curves for $T>T_C$ and $T<T_C$. b) In the modified Arrott plots, $|M|^{1/\beta}$ vs $|H/M|^{1/\gamma}$, the $M(H)$ data are linearized and evenly spaced in temperature.}
    \label{scaling}
\end{figure}

To analyze $M(H)$ data from single crystals, a critical scaling approach based on the Arrott-Noakes equation of state \cite{Arrott67} has been employed.  This approach does not assume mean-field behavior, and yields values of $T_{C}$, the ordered moment $M_0$ and the magnetic critical exponents $\beta$, $\gamma$, and $\delta$. The three exponents are defined in terms of $M$, $H$, and reduced temperature $t = \frac{(T - T_{C})}{T_{C}}$ by $M \sim t^{\beta}$ for $t<0$, $M \sim H^{1/\delta}$ for $t=0$, and $\frac{\partial M}{\partial H} = \chi \sim t^{-\gamma}$ for $t>0$. Only two exponents are independent, as $\delta-1 = \gamma/\beta$.  The self-consistent determination of their values requires agreement between three different analyses:  an initial estimate of  $\delta$ and $T_C$ made by identifying the isotherm with constant power-law behavior over the widest range of $H$; scaling in $|M|/|t|^{\beta}$ vs $H/|t|^{\delta\beta}$, i.e., the collapse of $M(H)$ data onto two diverging curves, for $t>0$ and $t<0$; and modified Arrott plots of $|M|^{1/\beta}$ vs $|H/M|^{1/\gamma}$.  Figure~\ref{scaling} exhibits the results of the scaling analysis for the $M(H)$ data measured on URu$_{1.70}$Re$_{0.30}$Si$_{2}$.  This analysis has been applied successfully to samples with $0.20 \leq x \leq 0.60$, which demonstrates the existence of FM order in this range of Re concentration \cite{Butch09}.

\begin{figure}[tbp]
    \begin{center}
    \rotatebox{0}{\includegraphics[width=8.5cm]{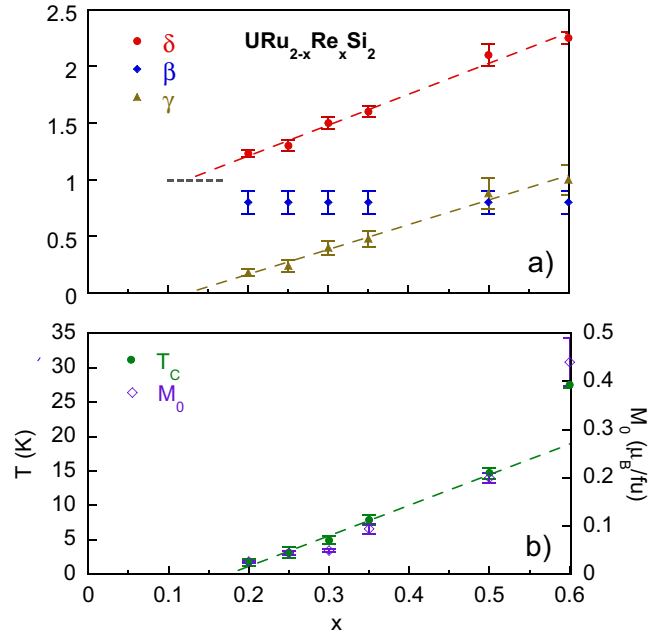}}
    \end{center}
    \caption{Re concentration dependence of magnetic critical exponents, critical temperature, and ordered moment of  URu$_{2-x}$Re$_{x}$Si$_{2}$. The $x$-dependence of $\delta$ and $\gamma$, and  $T_C$ and $M_0$ (for $x \leq 0.5$) is well-described by linear fits. The values of $\delta-1$, $\gamma$, $T_C$, and $M_0$ extrapolate to zero near $x=0.15$. Error bars denote the range of values that satisfy the scaling analysis. }
    \label{exponents}
\end{figure}

In Fig.~\ref{exponents}, it is apparent that the $x$-dependence of the exponents is linear for all $0.20 \leq x \leq 0.60$, but the $x$-dependence of $T_C$ and $M_0$ can be fit to a line only for $x \leq 0.50$.  In URu$_{2-x}$Re$_{x}$Si$_{2}$,  $T_{C}$, $M_{0}$, $\gamma$ and $(\delta-1)$ all extrapolate to $0$ at $x = 0.15\pm0.03$. This is the first documented example of such a trend.  The critical exponents deviate substantially from mean-field values and values describing classical FM transitions, where $\beta < 0.5$ and $\delta > 3$. Near the quantum phase transition, as $x$ decreases, $\delta \rightarrow 1$ as the $M(H)$ isotherms (at $T=T_C$) become less curved. Because $\beta$ remains constant, $\gamma \rightarrow 0$, implying a sharpening of the curvature of the divergent susceptibility at $T_C$.  While this behavior looks like a trend towards a first-order transition, note the simultaneous suppression of the order parameter $M_0$.

\begin{figure}[tbp]
    \begin{center}
    \rotatebox{0}{\includegraphics[width=8.5cm]{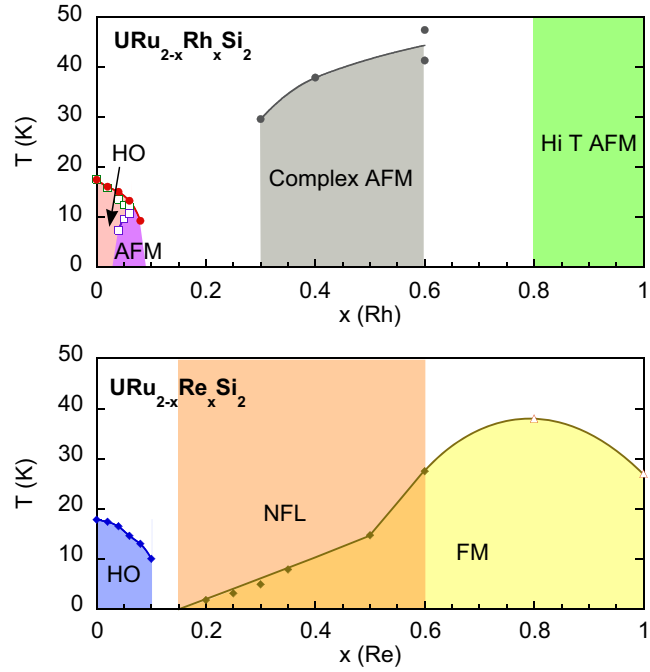}}
    \end{center}
    \caption{Comparison of the phase diagrams of URu$_{2-x}$Rh$_x$Si$_2$ and URu$_{2-x}$Re$_x$Si$_2$ up to 50~\% substitution. In URu$_{2-x}$Rh$_x$Si$_2$, the hidden order appears to be suppressed by $x=0.10$, although for $0.04 \leq x \leq 0.06$, HO gives way to larger-moment AFM order.  The complex AFM and high-$T$ AFM phases are characterized by local-moment order. Closed circles follow Ref.~\cite{Kawarazaki94} and open squares follow Ref.~\cite{Yokoyama04}. In URu$_{2-x}$Re$_x$Si$_2$, the hidden order persists to $x=0.10$.  Ferromagnetic order appears to emerge near $x \approx 0.15$, and $T_C$ increases with $x$ until $x=0.8$.  Structural heterogeneity sets in near $x=1$. Open triangles follow Ref.~\cite{Dalichaouch89}.  The range of NFL behavior was established in Refs.~\cite{Bauer05,Krishnamurthy08}.}
    \label{phasediagram}
\end{figure}

The Re concentration dependence of the HO and FM phases in URu$_{2-x}$Re$_x$Si$_2$ is shown in Fig.~\ref{phasediagram}.  Although the HO transition can only be determined for $x \leq 0.10$, the HO phase boundary does extrapolate to 0~K near $x \approx 0.15$, where it might meet the FM phase at a  multicritical point.  There is no direct evidence yet that the HO and FM phases meet, but it has been observed that AFM  correlations and magnetic excitations associated with the hidden order phase persist to $x=0.35$ \cite{Krishnamurthy08}.  Such correlations provide a plausible explanation for the unusual critical behavior of the FM ordered phase.  Another possible cause is a competition between magnetic order and the heavy fermion state in URu$_{2-x}$Re$_x$Si$_2$.  However, the details of this interplay are not certain, nor is it known whether itinerant FM arises from a light band or a heavy band in URu$_{2-x}$Re$_x$Si$_2$. FM Kondo lattice models for both light and heavy magnetic bands are consistent with the measured enhanced specific heat \cite{Perkins07,Yamamoto08}. Also shown in Fig.~\ref{phasediagram} is the range of $x$ over which NFL behavior has been observed \cite{Bauer05,Krishnamurthy08}, which appears to coincide with the FM ordered phase. There is some correlation between the NFL effects and the novel critical behavior: power-law exponents describing unconventional $T$-dependence of the magnetic susceptibility \cite{Bauer05} are in good agreement with values of $\gamma$ determined by the scaling analysis, implying that they actually describe the same phenomenon.

The overall phase diagram of URu$_{2-x}$Rh$_x$Si$_2$ \cite{Kawarazaki94,Yokoyama04} differs significantly from that of URu$_{2-x}$Re$_x$Si$_2$ (Fig.~\ref{phasediagram}), despite similar suppression of the HO phase at low $x$. Whereas URu$_{2-x}$Rh$_x$Si$_2$ appears to only support AFM order, the long-range magnetic order in URu$_{2-x}$Re$_x$Si$_2$ is FM. The temperature scales differ rather dramatically between the two phase diagrams, as well.  In particular, at high Rh concentrations, AFM order sets in at $\sim 180$~K, whereas in the Re phase diagram, $T_C$ reaches only $\sim 40$~K.  It is interesting that the intermediate AFM phase of URu$_{2-x}$Rh$_x$Si$_2$, characterized by a multi-$q$ domain structure \cite{Kawarazaki94}, persists over roughly the same range as the NFL behavior does in the Re case. The complex magnetic structure in URu$_{2-x}$Rh$_x$Si$_2$ and the AFM-like correlations in the FM phase of URu$_{2-x}$Re$_x$Si$_2$ \cite{Krishnamurthy08} suggest that in both cases, competing magnetic correlations in this range of $x$ play an important role in determining the magnetic structure.  An open question  is whether the HO phase in URu$_{2-x}$Re$_x$Si$_2$ gives way to an AFM phase, as it does in the Rh case \cite{Yokoyama04}.

\ack

We thank B. T. Yukich and T. A. Sayles for experimental assistance.  Sample preparation was supported by the U.S. Department of Energy (DOE) under Research Grant \# DE-FG02-04ER46105. Measurements where supported by the National Science Foundation under Grant No. 0802478.

\newpage

\end{document}